\documentclass[3p,times,procedia]{elsarticle}
\usepackage{nupha_ecrc}
\usepackage{graphicx,subfigure,latexsym,amssymb}
\usepackage{bm}
\usepackage{amsmath}

\newcommand{\be}{\begin{equation}}
\newcommand{\ee}{\end{equation}}
\newcommand{\bear}{\begin{eqnarray}}
\newcommand{\eear}{\end{eqnarray}}
\newcommand{\ba}{\begin{array}}
\newcommand{\ea}{\end{array}}

\def\({\left(}
\def\){\right)}

\volume{00}

\firstpage{1}

\journalname{Nuclear Physics A}

\runauth{S. Li}

\jid{nupha}

\jnltitlelogo{Nuclear Physics A}

\usepackage{amssymb}
\usepackage{amsthm}

\usepackage[figuresright]{rotating}

\begin{document}

\begin{frontmatter}


\title{Title\tnoteref{label1}}

\dochead{XXVIIIth International Conference on Ultrarelativistic Nucleus-Nucleus Collisions\\ (Quark Matter 2019)}

\title{Quantum Kinetic Equation for spin polarization of massive quarks from pQCD}

\author{Shiyong Li}
\address{Department of Physics, University of Illinois, Chicago, Illinois 60607, USA}



\begin{abstract}
We review the recent progress in formulating a quantum kinetic theory for the polarization of spin-$\frac{1}{2}$ massive quarks from the leading-log order of perturbative QCD (pQCD). 
\end{abstract}


\end{frontmatter}
\section{Introduction}
\label{}
In the early stage of the expansion of fireball created by heavy-ion collisions (HIC), QCD plasma is believed to be in its deconfined state, whose degrees of freedom are quarks and gluons. The spin of quasiparticle can be polarized by the magnetic field and the fluid vorticity in non-central HIC. Some part of spin polarization of the quarks and gluons in this phase is transferred to that of hadrons during hadronization, and is experimentally observed as the spin polarization of hadrons, such as the polarization of $\Lambda$ hyperons reported by the STAR collaboration\cite{STAR:2017ckg}. Some of the early studies discuss the conversion of angular momentum between fluid vorticity and spin angular momenta of quasiparticles through interactioins (see \cite{Liang:2004ph, Gao:2007bc, Becattini:2007sr}, and reference therein), yet the total angular momentum of the system is still conserved.

But little is known about the dynamical evolution of the conversion process, especially in the case when one goes beyond the equilibrium treatment where collision term must play a role. In a time-varying system in HIC, whether the spin polarization of quasiparticles is driven into or out of equilibrium depends on the relaxation time of the spin polarization due to QCD interactions which relax it to equilibrium, and the time variation of backgrounds such as vorticity (or magnetic field) which drive it off equilibrium. If the latter is much slower than the former, then the system follows closely the instantaneous equilibrium state. Otherwise, the system is driven significantly off equilibrium, so that the spin polarization should be determined by solving the dynamical evolution of spin polarization. On the other hand, within the framework of constituent quark model, the spin of $\Lambda$ hyperons is mainly carried by strange quarks \cite{Chiapparini:1991ap}. Therefore, the microscopic theory framework that describes the dynamical evolution of spin polarization of massive quarks is needed, which would be a prerequisite to study that of hadrons too. The free streaming Boltzmann equation of the spin-$\frac{1}{2}$ massive quarks has been studied (see \cite{Mueller:2019gjj, Weickgenannt:2019dks, Gao:2019znl, Hattori:2019ahi}), but the collision term is still missing.

In this contribution, we formulate the collision terms of dynamical evolution equation of the spin polarization of massive quarks in the leading-log order of pQCD \cite{Li:2019qkf}. By massive quarks we mean strange quarks or more massive quark species. Therefore, it is reasonable to justify that quark mass is of the order of hard scale, $m \geq m_D \sim gT$(g is the QCD coupling constant)\footnote{Working in this regime justifies the neglecting of quark-gluon conversion process at leading-log order due to the impossibility of soft quark exchange that must be the same species of the incoming massive quark.}. The similar study of relaxation time for spin polarization of strange quarks is provided in \cite{Kapusta:2019sad}. See \cite{Yang:2020hri} for new development.

\section{Time evolution of spin density matrix in Schwinger-Keldysh formalism}

\subsection{Spin density matrix}
The evolution equation of spin density matrix $\hat{\rho}$ of a massive spin-$\frac{1}{2}$ quark should satisfy the "Lindblad equation" 
\be
{\partial \hat\rho\over\partial t}=-{i\over\hbar}[H_{\rm eff},\hat\rho]\ -\Gamma\cdot\hat\rho,\label{eqS}
\ee
where the effective one-particle Hamiltonian $H_{\rm eff}$ in 2-dimensional spin space, from a phenomenology point of view, takes the form as $H_{\rm eff}=-{\hbar\over 2}\bm\sigma\cdot (\bm\omega+e\bm B)$. $\Gamma$ is the relaxation operator that we aim to study. In general, it can be expanded in terms of  small vorticity $\bm\omega$ and magnetic field $\bm B$ as $\Gamma=\Gamma_0+\Gamma_1(\bm\omega, \bm B)+\cdots$, where $\Gamma_0$, $ \Gamma_1(\omega, \bm B)$ are the zeroth, and first order in $\bm\omega$ or/and $\bm B$, respectively. As a first step, we present our result for the leading relaxation operator $\Gamma_0$, which describes how the initially polarized spin density matrix relaxes to the unpolarized state in the absence of vorticity and magnetic field, and the study of next order $\Gamma_1$ with the vorticity and magnetic effect is in progress. 

A convenient way to study the density matrix $\hat\rho$ in phase space $(\bm x,\bm p)$ is in the language of Schwinger-Keldysh contour. The position and momentum operators $(\bm x_{1(2)}, \bm p_{1(2)})$ evolve along the  forward or backward time contours(labels by 1 or 2). One can introduce "ra" variables where classical position and momentum are $\bm x_r={1\over 2}(\bm x_1+\bm x_2)$, $\bm p_r={1\over 2}(\bm p_1+\bm p_2)$, and quantum fluctuation of position and of momentum are $\bm x_a=\bm x_1-\bm x_2$, $\bm p_a=\bm p_1-\bm p_2$, respectively. Because of $[\bm x^i_{r},\bm p^j_{r}] =0$, it allows us to introduce density matrix in phase space as $\hat\rho(\bm x_r,\bm p_r)$. Due to that $\bm x_r$ is conjugate with $\bm p_a$, the density matrix in momentum space $\hat\rho(\bm p_r,\bm p_a)=\hat\rho(\bm p_1,\bm p_2)$ is related to the density matrix in phase space $\hat\rho(\bm x_r,\bm p_r)$ by a Fourier (or Wagner) transform
\be
\hat\rho(\bm x_r,\bm p_r)=\int {d^3\bm p_a\over (2\pi)^3}\,e^{i\bm p_a\cdot\bm x_r}\,\hat\rho(\bm p_r,\bm p_a)\,.\label{wigner}
\ee

Focus our discussion on the density matrix in diagonal momentum space, $\hat\rho(\bm p_1, \bm p_2) \sim (2\pi)^3 \delta(\bm p_1 -\bm p_2)\hat\rho(\bm p_1)$, but we still keep the full spin matrix $\hat \rho(\bm p)$ in the spin space. Or equivalently, It means that we work in the spatial homogeneity system limit, since $\bm p_a\sim \partial_{\bm x}$. The above consideration is justified as long as the spatial gradient is much smaller than the inverse of mean free path of QCD interaction $l_{mfp}^{-1}$.

The $2\times 2$ spin density matrix in momentum space is defined by
\be
\hat\rho(\bm p)= {1\over 2}f(\bm p)+\bm S(\bm p)\cdot\bm\sigma\,,\label{helicitybasis}
\ee

where $f(\bm p)$ is the particle number distribution and $\bm S(\bm p)$ is the spin polarization density, the integration of which gives us the total number of quarks and total spin polarization per unit volume.
\be
N=\int {d^3\bm p\over (2\pi)^3}\, f(\bm p)\,,\quad
\bm S=\int {d^3\bm p\over (2\pi)^3}\, \bm S(\bm p)\,.
\ee
 
\subsection{Time evolution of the spin density matrix}
The Hamiltonian in one-quark picture\footnote{This reduced one-quark description is justified either by the condition $m \geq T$ which results in the thermal suppression of Dirac-Fermi statistics or in the early stages of HIC where massive quarks are scare.} is the sum of the free kinetic energy $H_0$, and $H_I$ which is the QCD interaction of quarks with background gluon fields. The interaction Hamiltonian $H_I$ from the field theory Hamiltonian is:
\be
H_I=g\int d^3\bm x\, \bar\psi(\bm x)\gamma^\mu t^a \psi(\bm x) A^a_\mu(\bm x)\,,\label{interHamiltonian}
\ee
where $\psi(\bm x)$ is the quark field operator, and $A^a_\mu(\bm x)$ is the gluon field with color index $a$ ($t^a$ are the color generators). The one-particle interaction Hamiltonian can be obtained from (\ref{interHamiltonian}).

The time evolution of the density matrix $\hat\rho(t)$ in the Schwinger-Keldysh formalism is: 
\be
\hat\rho(t)=\langle U_1(t,t_0)\hat\rho(t_0) U^\dagger_2(t,t_0)\rangle_A\,,\label{timeevol}
\ee
where $U_{1,2}(t,t_0)={\cal P}e^{-i\int_{t_0}^t dt' \, H_{1,2}(t')}$ are the unitary
time evolution operators with Hamiltonian $H^{1(2)}$.

The above average $\langle\cdots\rangle_A$ involves the thermal correlation functions of gluon fields $A^{(1)}$ and $A^{(2)}$ in the Schwinger-Keldysh contours. The thermal average of one point function vanishes $\langle A^{1(2)}\rangle = 0$, whereas the thermal average of two-point functions include the one loop Hard Thermal Loop(HTL) gluon self-energy, the cutting of which represents the scatterings with the background thermal particles. Therefore, we have to expand (\ref{timeevol}) to the quadratic order of $H_I$. Recall that in the interaction picture: $H_I^{int}(t) = U_0^{\dagger}(t)H_I(t)U_0(t)$, where $U_0(t)=e^{-i H_0 t}$ is the free evolution. Then we have
\bear
\hat\rho(t) - U_0(t)\hat\rho(0)U_0^\dagger(t) &=& \int_0^{t} dt_1\int_0^{ t}dt_2 U_0( t)\langle H_I^{{\rm int}(1)}(t_1)\hat\rho(0) H_I^{{\rm int}(2)}(t_2)\rangle_A U_0^\dagger(t) \nonumber\\
&+&(-i)^2 U_0( t) \int_0^{ t}dt_1\int_0^{t_1}dt_1' \langle H_I^{{\rm int}(1)}(t_1)H_I^{{\rm int}(1)}(t_1')\rangle_A \hat\rho(0) U_0^\dagger(t) \nonumber\\
&+& (i)^2 U_0( t) \hat\rho(0) \int_0^{ t}dt_2\int_0^{t_2}dt_2' \langle H_I^{{\rm int}(2)}(t_2')H_I^{{\rm int}(2)}(t_2)\rangle_A U_0^\dagger( t)\,,\label{evol}
\eear

\begin{figure}[t]
 \centering
 \includegraphics[height=4cm,width=15cm]{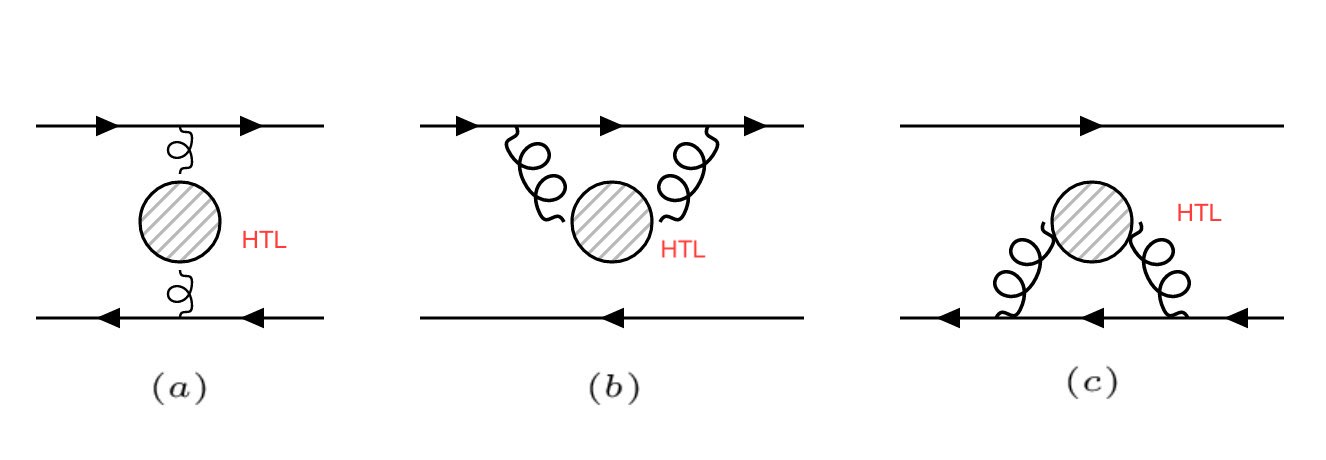}\caption{The "cross" diagram (a) corresponds to one $H_I$ from $U_1$ and one from $U_2$, and the two "self-energy" diagrams (b) and (c) coming from quadratic expansions in $H_I$ in each $U_1$ and $U_2$ in (\ref{evol}).  \label{fig1}} 
 \end{figure}
 
 The "cross" diagram in Fig.1.(a) is the loss term, which relaxes the spin polarization of massive quarks, whereas the "self-energy" diagram in Fig.1.(b) and (c) are the gain terms. The sum of all three diagrams preserves the total probability. Our result shows that the characteristic relaxation rate of spin polarization of massive quarks is of order $\alpha_s^2\log(1/\alpha_s)$. The $\log$ contribution is coming from the t-channal soft gluon exchange with momentum scale from the Debye screening mass $m_D \sim gT$ to hard scale $T$.
 
\section{Result and Summary}

We obtain a separate time evolution equation for the particle number distribution and the spin polarization density in the leading log order of $g^4\log(1/g)$:
\be
{\partial f(\bm p,t)\over\partial t}=C_2(F) {m_D^2 g^2\log(1/g)\over (4\pi)}\,{1\over 2p E_p}\,\Gamma_f\,,\quad
{\partial \bm S(\bm p,t)\over\partial t}=C_2(F){m_D^2 g^2\log(1/g)\over (4\pi)}\,{1\over 2p E_p}\,\bm\Gamma_S\,,
\ee

where $\Gamma_f$ and $\bm\Gamma_S$ are diffusion-like differential operators in momentum space. They are given by:
\bear
\frac{\Gamma_f}{2pE_p } &=& \bm \nabla_{p^i}\Bigl(T(\frac{3}{4} -\frac{E_p^2}{4p^2} + \frac{\eta_p m^4}{4p^3E_p})\bm \nabla_{p^i}f(\bm p)+ \bm p^i\frac{Tm^2}{4p^3E_p}(\eta_p +\frac{3E_p}{p}-\frac{3 \eta_p E_p^2}{p^2})\bm p \cdot \bm \nabla_p f(\bm p)  \nonumber\\
&+& \frac{\bm p^i}{2p^2} (E_p - \frac{\eta_p m^2}{p}) f(\bm p) \Bigr)
\eear

The above $\Gamma_f$ satisfies the nontrivial detailed balance condition, i.e. $\frac{\Gamma_f}{2pE_p }=0$ when $f(\bm p)$ takes a form of the Boltzmann distribution function $f^{\rm eq}(\bm p)=z e^{-E_p/T}$, where z is an arbitrary fugacity. It has been written as a total divergence so that the particle number is manifestly conserved. For $\bm \Gamma^i_S$, we have
\bear
\bm \Gamma^i_S &=& \left(2p + \frac{TE_p}{p} - \frac{ \eta_p m^2 T}{p^2}\right)\bm S^i(p) + \left(p T E_p - \frac{m^2 T E_p}{2p} + \frac{\eta_p m^4 T}{2p^2}\right)\bm\nabla_p^2 \bm S^i(\bm p) \nonumber\\
&+& \left(\frac{\eta_p m^2 T}{2p^2}\left(1-\frac{3E^2_p}{p^2}\right) + \frac{3m^2T E_p}{2p^3}\right)(\bm p\cdot\bm\nabla_p)^2\bm S^i(\bm p)\nonumber \\
&+&\frac{1}{p^2}\left(pE^2_p -\frac{3m^2TE_p}{2p} + \eta_p m^2\left(-E_p -\frac{T}{2}+\frac{3TE^2_p}{2p^2}\right)\right)(\bm p\cdot\bm\nabla_p)\bm S^i(\bm p)\nonumber \\
&+&2T \left(\eta_p\left(\frac{1}{2} -\frac{E^2_p}{p^2} + \frac{mE_p}{2p^2} + \frac{E^3_p}{2p^2(E_p + m)}\right) + \frac{E_p}{p} -\frac{m}{2p} -\frac{m^2}{2p(E_p + m)}\right)\left(\bm p^i(\bm\nabla_p\cdot\bm S(\bm p)) - \bm\nabla_p^i(\bm p\cdot\bm S(\bm p))\right)\nonumber \\
&-&\frac{T}{p^2}\left(\frac{E_p(E_p + 2m)}{p(E_p + m)} + \frac{\eta_p m E_p}{E_p+m}\left(-\frac{3E_p}{p^2}+\frac{1}{E_p +m}\right)\right)\bm p^i ({\bm p}\cdot\bm S(\bm p))\,.
\eear

Again one can check that $\frac{\bm \Gamma_s^i}{2pE_p}$ satisfies the nontrivial detailed balance condition in the massless limit when spin polarization density $\bm S(\bm p)$ takes a form of $\bm S(\bm p) = ze^{-\frac{p}{T}}\hat{\bm p}$. 

We have derived the time evolution equation for the spin polarization of spin-$\frac{1}{2}$ massive quarks. It can be further improved in the future by going beyond the spatial homogeneity limit, i.e. by including the spatial gradients of the density matrix, and considering the vorticity and magnetic field effect.

I thank Ho-Ung Yee for fruitful collaborations. This material is based upon work supported by the U.S. Department of Energy, Office of Science, Office of Nuclear Physics, with the grant No. DE-SC0018209 and within the framework of the Beam Energy Scan Theory (BEST) Topical Collaboration.










%

\end{document}